\documentstyle[manuscript,prd,aps]{revtex}
\begin{document}
\tighten
\draft
\title{DYNAMICAL STABILITY OF WITTEN RINGS.}
\author{Xavier MARTIN$^1$ and Patrick PETER$^{1,2}$}
\address{$^1$Departement d'Astrophysique Relativiste et de Cosmologie,\\
Observatoire de Paris-Meudon, UPR 176, CNRS, 92195 Meudon (France)\\
$^2$Department of Applied Mathematics and Theoretical Physics,\\
University of Cambridge, Silver street, Cambridge CB3 9EW, (UK)}
\preprint{DAMTP-R94/19,hep-ph/9405220}
\maketitle
\begin{abstract}
The dynamical stability of cosmic rings, or vortons, is investigated
for the particular equation of state given by the Witten bosonic
model. It is found that there exists a finite range of the state
parameter for which the vorton states are actually stable against
dynamical perturbations. Inclusion of the electromagnetic self action
into the equation of state slightly shrinks the stability region but
otherwise yields no qualitative difference. If the Witten bosonic
model represents a good approximation for more realistic string
models, then the cosmological vorton excess problem can only be solved
by assuming either that strings are formed at low energy scales or
that some quantum instability may develop at a sufficient rate.
\end{abstract}
\pacs{PACS numbers: 98.80.Cq, 11.27+d}

\section{Introduction}

Cosmic strings~\cite{kibble} have been proposed as seeds for large
scale structure formation~\cite{lss} and as a means to reproduce
temperature fluctuations in the cosmic microwave
background~\cite{cobe} if they appear at the Grand Unified (GUT) phase
transition. However, the underlying string theories in these
calculations assume structureless strings, i.e. the kind for which the
equation of state is that of Goto-Nambu (having their energy per unit
length $U$ equal to their tension $T$, both being strictly constant
along the string worldsheet), so the question arises to what extent
these conclusions are still valid in more complicated cases such as
those proposed by Witten~\cite{witten} where a charged condensate is
formed in the string's core, thereby inducing a current and thus
modifying the equation of state into a nondegenerate one. It has been
argued~\cite{vorton,ring} that in the latter case, equilibrium
configurations (called vortons~\cite{vorton} or rings~\cite{ring})
might exist, that would not radiate all their energy in the form of
gravitational waves (which is the case for Goto-Nambu strings), and
therefore contribute non negligibly to the overall matter density in
the Universe. Estimates of the corresponding vorton distribution led
Davis and Shellard~\cite{vorton} and Carter~\cite{ring} to the
conclusion that the stability of such states would imply a huge
matter-density remnant, so that current-carrying strings appearing at
the GUT phase transition would induce $\sim 10^{20}$ times the
critical density, and that a premature collapse of the Universe would
be avoided only if they are lightweight (being produced at energy
scales at most comparable to that of the electroweak phase
transition).

There are in fact two necessary ingredients for this overdensity
problem. The first concerns the currents in the strings, since this is
an essential requirement for the equilibrium configurations to exist:
rings are centrifugally supported string loops and thus are only
defined when the equation of state is nondegenerate so that there
exists a prefered frame in which rotation makes sense. This issue has
been addressed in a particle physics framework with the following
conclusion. If the string forming GUT theory is to have as a low
energy limit the standard electroweak model (a ``naturalness''
requirement), then the string forming fields cannot be arbitrarily
decoupled from electromagnetism, and currents will appear in strings,
either through quantum tunnelling~\cite{low-mass}, or by spontaneous
current generation~\cite{scg}. This can in fact be traced back to the
nonabelian nature of the GUT model being used: since the string
forming Higgs field expectation value vanishes in the string's core,
some charged gauge vectors are massless there, and induce an
electromagnetic instability~\cite{scg} or
metastability~\cite{low-mass}, depending on the coupling constant
values. In both cases, currents appear in strings as a generic
feature, so the problem remains. Moreover, it has also been
shown~\cite{low-mass} that even in these realistic (and therefore more
complicated) models, the equation of state can be well approximated by
that of the Witten bosonic model, which was recently calculated in
detail~\cite{state,neutral,enon0}. This is the reason why this
particular equation of state is used in the present work.

The second necessary requirement to have a string dominated Universe
is that the equilibrium configurations be classically stable, at least
dynamically. This is not an easily verified assumption, and the
purpose of this work is precisely to investigate it in the framework
of the Witten bosonic model, both in the neutral limit~\cite{neutral}
case where the electromagnetic coupling is made to vanish, and in the
charge coupled case, by implementing the actual value of this coupling
constant on the equation of state~\cite{enon0}, including back
reaction on the constituant fields.  Since this work is primarily
concerned with dynamical stability, we have not included the
electromagnetic corrections either to the the equilibrium condition or
to the stability constraint, these (expected stabilising) corrections
being of a higher order and left for further examination~\cite{ecorr}.

This work is arranged as follow. In a first part, we recall the basic
dynamical equations for a string loop, and the equilibrium condition,
as well as the stability constraints~\cite{martin}. This is done
assuming an underlying equation of state, or, more precisely, by
assuming the characteristic perturbation velocities to be given. These
velocities correspond to transverse and soundlike perturbations, being
expressible respectively as $c_T^2 = T/U$ and $c_L^2 = -dT/dU$. Then,
a brief summary of the characteristic features of the Witten bosonic
model is given and we ultimately apply the stability calculations for
this model to show that a finite range of currents actually yields
stable states.  Since it turns out that this range is for low values
of the current, or for nearly lightlike currents, we argue that most
vortons are in fact stable, so that, unless quantum radiation provide
some efficient way to destabilise them, they can only have been
created at a low energy phase transition.

\section{Stability of a ring configuration}

In order to calculate the dynamical evolution of a cosmic string, we
assume it to be infinitely thin and characterize the corresponding
two-dimensional worldsheet by a set of two orthonormal tangent vectors
$u^\mu$ and $v^\mu$, respectively timelike and spacelike. These
vectors allow an easy definition of the first and second fundamental
tensors of the string's worldsheet, namely~\cite{meca-ring,meca-ring2}
\begin{equation} \eta ^{\mu \nu } \equiv -u^{\mu }u^{\nu } +v^{\mu
}v^{\nu }, \ \ \ K_{\mu \nu } {^{\rho }} \equiv \eta
^{\sigma } {_{\mu }} \eta^\lambda {_{\nu}}
\nabla _{\lambda } \eta ^{\rho } {_{\sigma }},\label{K}\end{equation}
the latter satisfying the Weingarten identity
\begin{equation} K_{[\mu \nu ]} {^{\rho }}=0 ,\label{geom}\end{equation}
which is the integrability condition for the existence of a worldsheet
containing $u^\mu$ and $v^\mu$ as tangent vectors.

Since the string is considered as infinitely thin, its stress-energy
tensor $T^{\mu\nu}$ is a distribution defined on the string worldsheet
only, and therefore involves in principle singular $\delta
-$functions. However, use of such distributions can be
avoided~\cite{meca-ring2} by working with a surface stress-energy
tensor $\widetilde{T} ^{\mu\nu }$, depending on the internal
coordinates of the worldsheet, $\tau$ and $\sigma$ say, such
that the corresponding distribution valued tensor $T^{\mu\nu}$ can, if
needed, be obtained from the formula
\begin{equation} T^{\mu \nu }(x)=(-g)^{-1/2} \int d\! S_{2}
\widetilde{T} ^{\mu \nu
} (\tau ,\sigma )\delta ^{4} (x- x(\tau ,\sigma ))
,\label{T}\end{equation} where $x(\tau ,\sigma )$ is a generic point
of the worldsheet, $d\!  S_{2}$ the surface measure and $g$ is the
determinant of the metric $g^{\mu \nu }$ of the surrounding four
dimensional space-time (which is supposed to be flat in the following
analysis). Conservation of the stress-energy tensor $T^{\mu \nu }$
then implies that the surface stress tensor satisfies a corresponding
worldsheet conservation law having the form~\cite{meca-ring}
\begin{equation}  \eta ^\lambda {_{\nu}} \nabla _\lambda \widetilde
T^{\mu \nu} = 0.\label{dyn}\end{equation}
This is the dynamical part of the string motion. It can be further
simplified by choosing the frame ($u^\mu$, $v^\mu$) used in
Eq.~(\ref{K}) to be identified with the frame in which $\widetilde
{\bf T}$ is diagonal, with eigenvalues $U$ and $T$ respectively the
energy per unit length and tension, in the form
\begin{equation} \widetilde{T} ^{\mu \nu }=Uu^{\mu }u^{\nu }-Tv^{\mu
}v^{\nu }. \label{Ttilde}\end{equation}

It is convenient for computational purposes to start with $u^\mu$ and
$v^\mu$ as basic independent variables whose integration determines
the worldsheet as a secondary construct. In such an
approch~\cite{martin}, the Weingarten condition~(\ref{geom}) is not
satisfied automatically but must be included as an additional
dynamical equation together with Eq.~(\ref{dyn}). Subject to provision
of the equation of state (see next section), Eqs.~(\ref{geom})
and~(\ref{dyn}) provide a complete description of the string motion,
expressible in terms of the transverse perturbations and the
longitudinal group velocities
\begin{equation} c_{T}=\sqrt{\frac{T}{U}} \mbox{ , }
c_{L}=\sqrt{-\frac{dT}{dU}} ,\end{equation}
as the following system:
\begin{eqnarray} K_{[ \mu \nu ]} {^{\rho }}=0\Leftrightarrow &
\perp _{\mu \nu} (u^{\rho }\nabla _{\rho }
(v^{\nu })-v^{\rho }\nabla _{\rho } (u^{\nu }))= & 0, \label{eq1}\\
\perp _{\mu \nu} \overline{\nabla }_{\rho } \widetilde T^{\rho \nu }= &
\perp _{\mu \nu} (u^{\rho }\nabla _{\rho } (u^{\nu })-
c_{T} ^{2} v^{\rho } \nabla_{\rho } (v^{\nu }))= & 0, \label{eq2}\\
u_{\nu }\overline{\nabla }_{\rho } \widetilde T^{\rho \nu }= &
Au^{\rho }\nabla _{\rho}(c_{T}^{2})-(1-c_{T}^{2}) u_{\nu }v^{\rho
}\nabla_{\rho }(v^{\nu })= & 0, \label{eq3}\\ v_{\nu }\overline{\nabla
}_{\rho } \widetilde T^{\rho \nu }= &-Ac_{L}^{2} v^{\rho }\nabla
_{\rho}(c_{T}^{2})+(1-c_{T}^{2})u_{\nu }u^{\rho }
\nabla_{\rho } (v^{\nu })= & 0, \label{eq4}\end{eqnarray}
where $\perp _{\mu \nu} =g_{\mu \nu}-\eta _{\mu \nu}$ is the
orthogonal projection operator and $A=-(c_T^2+c_L^2)^{-1}$. The
unknown quantities~\cite{martin2} associated with this system can be
chosen as the five independent components of the tangent vectors
$u^{\mu }$ and $v^{\mu }$ and $c_T^2 $, while $c_L^2$ is given a
priori as a function of $c_T^2$ by the equation of state.

Let us now more specifically turn to the case of a circular rotating
string with radius $R$, angular speed $\Omega $ and running velocity
$v=R\Omega $, for which one can express the eigenvectors in
cylindrical coordinates $(t,\rho ,\theta ,z)$ as
\begin{equation}
u^{\mu }=\gamma (1,0,v,0) \ \ \ \mbox{ and } \ \ \ v^{\mu }=\gamma (v,0,1,0),
\label{ringconf} \end{equation}
where $\gamma =1/\sqrt{1-v^2}$ is the Lorentz factor associated to the
running velocity $v$. It may be shown~\cite{ring,martin}, that the
ring configuration~(\ref{ringconf}) automatically satisfies the system
of Eqs.~(\ref{eq1})-(\ref{eq4}), except for Eq.~(\ref{eq2}) which
provides the simple relation (again, not taking into account the small
long-range electromagnetic back reaction~\cite{lett})
\begin{equation} v=c_{T} .\label{equil}\end{equation}
Configurations for which Eq.~(\ref{equil}) hold are the equilibrium
states (vortons or rings) whose condition of stability is summarized
below. This relation is in fact more general in the sense that it must
hold also for any non straight equilibrium state with a static Killing
vector~\cite{martin}.

We now consider the perturbations of the equilibrium
state~(\ref{equil}).  Without loss of generality, and thanks to the
symmetries of this configuration, the perturbed quantities can be
taken in the form of plane waves with pulsation $\omega $ and integer
angular momentum $n$, so that, in particular, the perturbed parameters
describing the string's location read as
\begin{eqnarray} \rho & = & R+\delta \! R \mbox{e} ^{i(\omega
t-n\theta )} ,\\ z & = & \delta \! z \mbox{e} ^{i(\omega t-n\theta )}
.\end{eqnarray}
with $\delta \! R$ and $\delta \! z$ constant amplitudes small compared
to the unperturbed value $R$. The 6 perturbed independent
quantities\cite{martin2} in this particular case are chosen as $\delta
\!  u^{\rho}$, $\delta \! u^{\theta}$, $\delta \! u^{z}$, $\delta \!
v^{\rho}$, $\delta \! v^{z}$ and $\delta \! c_{T}^2$, out of which
one can reconstruct the full perturbed ring, and in particular $\delta
\! R$ and $\delta \! z$, this reconstruction being consistent only in
the case where the geometrical Eq.~(\ref{geom}) is satisfied.

The equations~(\ref{eq1})~--~(\ref{eq4}) give a linear homogeneous
system of six equations with six unknowns in which the azimuthal
perturbations as exemplified by $\delta \! u^{z} $ and $\delta \!
v^{z} $ decouple from the equatorial ones consisting of the other
unknowns $\delta \! u^{\rho} $, $\delta \! u^{\theta} $, $\delta \!
v^{\rho} $ and $\delta \! c_{T}^2$. The azimuthal part of this system
yields the modes~\cite{martin2}
\begin{equation} \omega \in \left\{ 0,{2nv\over R (1+v^2)} \right\}
\label{azim}\end{equation}
which are all stable (their imaginary part vanishing). Therefore, we
shall only be concerned by the equatorial modes, given~\cite{martin}
as the solutions of the third degree polynomial in $\sigma \equiv
\omega /\Omega$
\begin{eqnarray} v^2 (1+v^2)(1-c^2 v^2) \sigma ^3 + 2nv^2 [c^2-v^2
-2(1-c^2 v^2)]\sigma^2 + & &\nonumber\\
\null [4v^2(1-c^2)(n^2-1)-(1+v^2)(c^2 -v^2)(n^2+1)]\sigma
+2n(c^2-v^2)(n^2-1) & = & 0, \label{poly3}\end{eqnarray} with $c\equiv
c_L$, and where the static mode $\sigma =0$ has been implicitely
extracted out. This equation might have complex roots, in which case
the corresponding state will be unstable, with characteristic
life-time $\tau ^{-1} = |{\cal I}$m$\, \omega |$. A previous
analysis~\cite{martin2}, using explicit solutions of Eq.~(\ref{poly3})
given by the Cardan formulae, exhibited the stability and instability
regions in the square ($c_L^2,c_T^2$) as well as provided analytic
expressions for the imaginary part of the modes in the latter case.
It was found that while approaching the corner $c_L^2=c_T^2=1$ (where
the Witten model is located), the zones where instabilities might develop
turn into vanishingly thin surfaces so that almost any equation of
state must here cross stable zones.  This observation was the
starting point for the following closer examination of the Witten
model in this context.

\section{The Witten bosonic model}

The most simple model that leads to superconducting cosmic strings
consists in the abelian Higgs model in which a Higgs field $\Phi$
breaks a U(1) symmetry by means of a nonvanishing vacuum expectation
value $\langle |\Phi |\rangle = \eta$, thereby giving a mass to a
gauge vector boson $B_\mu$, coupled to a simplified representation of
electromagnetism in which a charged scalar field $\Sigma$
(representing some sort of average of all the various possible fields
involved in the underlying theory) is coupled to the photon $A_\mu$.
The general Lagrangian density that describes such fields is
\begin{equation} {\cal L} = -{1\over 2} |D_\mu \Phi |^2 -{1\over 2}
|D_\mu \Sigma |^2 - {1\over 16\pi} H_{\mu\nu}{^2} - {1\over 16\pi}
F_{\mu\nu}{^2} - {\cal V}(\Phi ,\Sigma ) ,\label{lag}\end{equation}
where the covariant derivatives are defined through
\begin{equation} D_\mu \Phi \equiv (\partial_\mu +iqB_\mu )\Phi
, \ \ \ D_\mu \Sigma \equiv (\partial_\mu +ieA_\mu )\Sigma ,\end{equation}
the Maxwell tensors are
\begin{equation} H_{\mu\nu} = \partial _{[\mu } B_{\nu ]}, \ \
\ F_{\mu\nu} = \partial _{[\mu } A_{\nu ]},\end{equation}
and the interaction potential can be taken as
\begin{equation} {\cal V} (\Phi ,\Sigma ) = {\lambda_\phi\over 8}
(|\Phi |^2 - \eta ^2)^2 + f (|\Phi |^2 -\eta^2)|\Sigma |^2 +
{\lambda_\sigma\over 4} |\Sigma |^4 + {m_\sigma ^2\over 2} |\Sigma |^2
.\end{equation} This model allows vortex solutions with a
$\Sigma$-condensate responsible for a change in the equation of state
as we now recall.

In order to study the stability of ring configurations, it is
necessary either to calculate the full field equations in a circular
vortex configuration, or, assuming the thin string approximation to
hold, to assume the string locally straight for the fields, calculate
the various integrated (on a transverse plane) components of the
stress energy tensor, and plug these back into Eq.~(\ref{T}). These
values can afterwards be used in the previouly developed formalism, so
it is this approach that we shall follow here.  Thus, we shall
consider a portion of straight string, which we choose to be aligned
with a coordinate axis $z$, and, working in cylindrical coordinates,
we study a Nielsen-Olesen~\cite{NO} vortex solution of the form
\begin{equation} \Phi = \varphi (r) \exp (iN\theta ), \ \ \
\Sigma = \sigma (r) \exp [i\psi (z,t)] ,\end{equation}
for some integer winding number $N$, $\psi $ being possibly restricted
to the simple form
\begin{equation} \psi (z,t) = a z - b t .\end{equation}
It can be shown~\cite{neutral,enon0} that the only arbitrary parameter
for a string in this model (assuming the underlying coupling constants
to be given) is the phase gradient
\begin{equation} \nu = \pm \sqrt{|\partial _\mu \psi \partial^\mu \psi
|}, \end{equation}
where the $+$ or $-$ sign must be chosen according to whether the
conserved electromagnetic current
\begin{equation} {\cal J}^\mu \equiv {1\over e}{\delta {\cal L}
\over \delta A_\mu} = \sigma ^2 (\nabla ^\mu \psi +eA^\mu
)\label{curr}\end{equation} is respectively spacelike or timelike. A
very important point concerning this current, and the corresponding
existence of centrifugally supported equilibrium states, is that it is
defined even in the so-called neutral limit~\cite{neutral} where the
electromagnetic coupling constant $e$ is made to vanish. This model
therefore accounts in particular for neutral-carrier condensates in
cosmic strings, and allows an easy recognition of the fact that the
long range feature of electromagnetism is essentially irrelevant in
ring dynamics. On the figures explained below, curves have been
plotted using both $e=0$ and a large value for this parameter
producing only minor {\it quantitative} corrections (exagerated
on the figures).

The other conserved quantity that is needed is the stress energy
tensor
\begin{equation} T^\mu_\nu = -2g^{\mu\alpha}{\delta {\cal L}\over
\delta g^{\alpha\nu}}+\delta ^\mu_\nu {\cal L}, \end{equation}
which yields, in the notation of the previous section, Eq.~(\ref{Ttilde})
\begin{equation} U = \widetilde T^{tt} = 2\pi \int r\,dr\,T^{tt}
\end{equation}
and
\begin{equation} T = -\widetilde T^{zz} = -2\pi \int r\,dr\,T^{zz}
.\end{equation}

The procedure now goes as follows: to each value of the state
parameter $\nu$ corresponds a (numerically computed) field
configuration which can be integrated to yield $U$ and $T$, with which
we calculate the velocities $c_T$ and $c_L$. (More details concerning
these computations can be found in particular in
Refs.~\cite{neutral,enon0}.) A characteristic result is shown on
Fig.~1 on which are plotted these velocities as functions of the state
parameter $\nu$ expressed in units of the mass of the current-carrier
$\Sigma$, in the neutral limit (dashed curves) and in the case where
the full electromagnetic self action has been taken into
account (full curves). Again, it should be stressed that the
corrections resulting from the inclusion of the electromagnetic
coupling constant $e$ affect our results only in a quantitave way, the
perturbation velocity plots being almost indistinguishable in most of
the parameter space.

\section{Results and Conclusions}

The Witten bosonic model has several underlying parameters, all of
which are supposed to be fixed by the underlying string-forming
microscopic theory. For each particular string, there is only one that
remains independent, namely the squared phase gradient $\nu ^2$ of the
current-carrier condensate, called the state parameter, whose sign
reflects the timelike or spacelike nature of the superconducting
current, and whose amplitude gives, in a nontrivial way, the amplitude
of the corresponding current, and the degeneracy of the stress-energy
tensor.  For various values of the underlying parameters, we have
derived the variations of the energy per unit length $U$ and the
tension $T$ with the state parameter, enabling us to calculate the
actual values, in these models, of the perturbation velocities
$c_T{^2} =T/U$ and $c_L{^2} = -dT/dU$, as illustrated on Fig.~1, both
for vanishing electromagnetic coupling constant and with inclusion of
the full self action on the string-forming fields. We have then used
these values to question the stability of ringlike configurations
(vortons) against azimuthal and equatorial perturbations, and we found
that generic results, meaning ones roughly independently of the values
of the underlying parameters (coupling constants $\lambda_\phi$,
$\lambda_\sigma$, $q$, $e$, masses, $\cdots$), could be drawn as
exemplified in Figs.~1 and 2.

The first point, as was already emphasized in previous work (but whose
implications had not yet been worked out in full detail), is that the
velocity of transverse perturbation always exceeds that of
longitudinal perturbations, even for very low current values, so that
the first order approximation for the evolution of a superconducting
string, namely that for which an effective action of the form
\begin{equation} S = \hbox{const} \int d^2 \! s (\nabla_\mu \psi +
eA_\mu )^2 ,\label{S1}\end{equation} where the transverse degrees of
freedom have been integrated out, is in fact unsatisfactory as soon as
one needs to consider derivatives of the equation of state. For such
purposes, the alternative action~\cite{meca-ring2}
\begin{equation} S = \hbox{const} \int d^2 \! s \sqrt{m^2+
(\nabla_\mu \psi + eA_\mu )^2 } ,\label{S2}\end{equation} with $m$ a
constant with the dimension of mass, should preferentially be used
since it reproduces most of the actual features of the Witten model,
and besides, is completely integrable (in the sense that analytic
solutions for the string worldsheet can be explicitly
constructed~\cite{warmcs,frolov}) and thus leads to much simpler
equations of motion. This may be understood by stating that the
crucial contribution for the derivatives involved in $c_L^2$ is given
by the fourth order term in the phase gradient, assumed to be
identically zero in the action~(\ref{S1}), though not in
Eq.~(\ref{S2}). More precisely, it can be seen that Eq.~(\ref{S1})
implies $c_L=1$, whereas in fact, in the Witten model, it is not only
less that the speed of light, but also less than $c_T$.  In that
sense, Eq.~(\ref{S2}) gives the better approximation $c_L=c_T$,
although stability considerations do not apply since this relation has
been shown~\cite{martin} to imply absolute stability against any
dynamical perturbations. For currents small enough, and when the
internal degrees of freedom of the strings are neglected, the
approximation~(\ref{S1}) can still be used for some purposes, but only
in a more restricted range of applications than previously thought.

Now the problem is that if the longitudinal velocity had been greater
than the transverse one, the ring dynamical stability issue could have
been addressed far more easily since in this region of the $(c_L,c_T)$
plane, circular configurations are always stable. However, it is less
obvious what will happen in the Witten string case: the regions of
instability in the $(c_L,c_T)$ plane reduce to lines close to the $c_L
= c_T =1$ edge where the most interesting part of the Witten-model's
equation of state is. Therefore, although it is not possible to show
it explicitely on a plot (the number of unstable zones is in principle
infinite as one approaches the $c_T=1$ line and it was not possible to
draw a clear graph), it appears that the equation of state will cross
various stable and unstable regions as it goes away from the stable
point $c_L=c_T=1$ (including in particular the null current case),
before ultimately reaching the unstable region.  This can be shown
alternatively as on Fig.~2 where the characteristic inverse life-time
of the configuration, expressed in units of the angular velocity
$\Omega$ of the ring, is calculated [as the larger imaginary part of
the solutions of Eq.~(\ref{poly3})] as a function of the state
parameter. This figure shows explicitely the range of phase gradient
where the corresponding ring states are stable. Various remarks need
to be made. First, the electromagnetic correction to the equation of
state yields a small correction to the actual size of the stability
region, but does not modify the result qualitatively. It should be
emphasized that for a realistic underlying field theory in which the
mass $M_\phi$ of the string forming Higgs particle is expected to be
much larger than $m_\sigma$, the relevant coupling constant would be
so small~\cite{enon0,nospring}
\begin{equation} e^2 (m_\sigma / M_\phi )^2 \sim 10^{-6}
\end{equation}
that its effect would be imperceptible. In order to obtain an effect
large enough to be visible on the figure, an artificially exagerated
value $e^2 (m_\sigma / M_\phi )^2 =0.1$ has therefore been used.
Therefore, one important conclusion that can again be drawn from these
results is that electromagnetic corrections to the field equations in
the string core do not significantly modify the actual dynamics of a
string, so that use of the neutral limit model~\cite{neutral} is
justified.

The last point we want to emphasize concerns the vorton excess
problem. As the equilibrium condition happens to be reached primarily
for quasinull currents~\cite{vorton}, for which $\nu$ is very small,
it can be argued, on the basis of our results, that most of these
vortons are in fact dynamically stable since this is precisely the
region where stability actually occurs. So our final conclusion is
that if the Witten bosonic model is a good approximation for
describing superconducting cosmic strings, then any string-forming
theory will produce dynamically stable vortons which can lead to an
overdensity in the Universe unless the phase transition happens at low
enough energy (estimated of order 10~TeV~\cite{ring}), or if quantum
instabilities can develop with sufficient rate. This can also be
understood as a constraint on the scheme of symmetry breaking in any
string forming GUT model or on the values of its microscopic
parameters.

\section*{acknowledgments}

We wish to thank B.~Carter and E.~P.~S.~Shellard for many stimulating
conversations. P.~P. was supported by SERC grant \#~15091-AOZ-L9.

\begin{figure}
\caption{Plot of the transverse and longitudinal velocities  $c_T^2 =
T/U$ and $c_L^2 = -dT/dU$ as functions of the phase gradient $\nu ^2=
|\partial _\mu \psi \partial^\mu \psi |$ of the current
carrier (in units of the inverse current-carrier mass). Dashed lines
are for the neutral limit with vanishing electromagnetic coupling
constant, whereas solid lines include the full electromagnetic back
reaction on the underlying fields, assuming an unreasonably large
value for the coupling constant $e$ to enhance the effect.}
\end{figure}

\begin{figure}
\caption{The inverse life-time $\tau ^{-1} = |{\cal I}$m$\, \omega |$
given by the complex roots of Eq.~(16) for the Witten bosonic model,
in units of the angular velocity $\Omega$ of the ring. A
finite region of the state parameter space has an infinite life-time
(i.e. yield stable rings), for small or quasi-lightlike currents, and
this is a generic feature in this model.}
\end{figure}

\end{document}